\documentclass[aps,graphicx,twocolumn,prd,showpacs]{revtex4-1}
\usepackage{graphicx}
\usepackage{dcolumn}
\usepackage{bm}
\begin{document}

\title{Pion form factor in the range $-10\mbox{ GeV}^2\leq s\leq1$ GeV$^2$.}
\author{N.~N.~Achasov}
\email[]{achasov@math.nsc.ru} \affiliation{Laboratory of
Theoretical Physics, S.~L.~Sobolev Institute for Mathematics,
630090, Novosibirsk, Russian Federation
}%
\author{A.~A.~Kozhevnikov}
\email[]{kozhev@math.nsc.ru} \affiliation{Laboratory of
Theoretical Physics, S.~L.~Sobolev Institute for Mathematics, and
Novosibirsk State University, 630090, Novosibirsk, Russian
Federation}

\date{\today}

\begin{abstract}
Based on the field-theory-inspired approach, a new expression for
the pion form factor $F_\pi$ is proposed. It takes into account
the pseudoscalar meson loops $\pi^+\pi^-$ and $K\bar K$ and the
mixing of $\rho(770)$ with heavier $\rho(1450)$ and $\rho(1700)$
resonances. The expression possesses correct analytical properties
and describes the data in the wide  range of the energy squared
$-10\mbox{ GeV}^2\leq s\leq1$ GeV$^2$ without introducing the
phenomenological  Blatt -- Weisskopf range parameter $R_\pi$.
\end{abstract}

\maketitle

The electromagnetic form factor of the pion $F_\pi$ is an
important characteristic of the low energy phenomena in particle
physics related with the hadronic properties of the
electromagnetic current in the theoretical scheme of the vector
dominance model \cite{Sak,gell61,KLZ}. There are a number of
expressions for this quantity used in the analysis of experimental
data in the time-like range $s>0$. Hereafter $s$ is the
center-of-mass-energy squared. The simplest approximate vector
dominance model expression based on the effective $\gamma-\rho$
coupling $\propto\rho_\mu A_\mu$ \cite{KLZ},
\begin{equation}
F_\pi(s)=\frac{m^2_\rho g_{\rho\pi\pi}/g_\rho}
{m^2_\rho-s-i\sqrt{s}\Gamma_{\rho\pi\pi}(s)},\label{Fpisingle}\end{equation}
does not possess the correct analytical properties upon the
continuation to the unphysical region $0\leq s<4m^2_\pi$ and
further to the spacelike region $s\leq0$, nor does it takes into
account the mixing of the isovector $\rho$-like resonances. Since,
phenomenologically \cite{pdg},
\begin{equation}
\frac{g_{\rho\pi\pi}}{g_\rho}=\left(\frac{9m_\rho\Gamma_{\rho\pi\pi}
\Gamma_{\rho ee}}{2\alpha^2q^3_{\pi}}\right)^{1/2}\approx1.20
\label{normFpi}
\end{equation}
the correct normalization $F_\pi(0)=1$ is satisfied by
Eq.~(\ref{Fpisingle}) only approximately, provided
$\Gamma_{\rho\pi\pi}(s)$ vanishes at $s\leq4m^2_\pi$. Hereafter,
$\alpha=1/137$ stands for the fine structure constant. The formula
of Gounaris and Sakurai \cite{GS} respects the normalization
condition and has correct analytical properties. However, being
based on some sort of effective radius approximation for the
single $\rho(770)$ resonance, it is not suited for taking into
account the mixing of $\rho(770)$ with heavier isovector mesons.
The expression  based on the gauge invariant $\gamma-\rho$
coupling $\propto\rho_{\mu\nu}F_{\mu\nu}$,
\begin{equation}
F_\pi(s)=1+\frac{sg_{\rho\pi\pi}/g_\rho}
{m^2_\rho-s-i\sqrt{s}\Gamma_{\rho\pi\pi}(s)},\label{Fpigau}\end{equation}
respects the correct normalization, but does not have correct
analytical properties  and breaks unitarity. The  earlier
expression \cite{ach97} for $F_\pi$  takes into account the strong
isovector mixing, but the normalization is satisfied within the
accuracy 20$\%$.

On the other hand, since the pioneer works \cite{rad,efr}, the
space-like region is considered as the test ground of the
perturbative QCD predictions for $F_\pi$ addressing, in
particular, the issues such as where the QCD asymptotic starts.

In the present work  the expression  for the pion form factor is
obtained which has correct analytical properties  in both the
space-like and time-like domains and takes into account the mixing
of $\rho(770)$ with the heavier resonances $\rho(1450)$ and
$\rho(1700)$. Remarkably, it does require the phenomenological
Blatt -- Weisskopf range factor $R_\pi$. By restricting the
consideration to the pseudoscalar meson loops $PP=\pi^+\pi^-$,
$K\bar K$ admitting the analytical treatment and valid at energies
below 1 GeV, the new expression is found and compared with the
existing data on $F_\pi$ collected with the detectors SND
\cite{snd} CMD-2 \cite{cmd}, KLOE \cite{kloe}, and BaBaR
\cite{babar} at $4m^2_\pi\leq s\leq1$ GeV$^2$. With the resonance
parameters found in this region the continuation to the region
$-10\mbox{ GeV}^2\leq s\leq0$ is made and compared with the
existing experimental data \cite{amendolia,bebek,horn,tadev}.

The new expression for the pion form factor is represented in the
form
\begin{eqnarray}
F_\pi(s)&=&(g_{\gamma\rho_1},g_{\gamma\rho_2},g_{\gamma\rho_3})G^{-1}\left(%
\begin{array}{c}
  g_{\rho_1\pi\pi} \\
  g_{\rho_2\pi\pi} \\
  g_{\rho_3\pi\pi} \\
\end{array}%
\right)+\nonumber\\&&
\frac{g_{\gamma\omega}\Pi_{\rho_1\omega}}{D_\omega\Delta}\left(g_{11}g_{\rho_1\pi\pi}+
g_{12}g_{\rho_2\pi\pi}+\right.\nonumber\\&&\left.
g_{13}g_{\rho_3\pi\pi}\right).\label{Fpi}\end{eqnarray} It takes
into account both the strong isovector
$\rho(770)-\rho(1450)-\rho(1700)$ mixing and the small
$\rho(770)-\omega(782)$ one,  automatically respects the current
conservation condition $F_\pi(0)=1$, and possesses correct
analytical properties over entire $s$ plane. The notations are as
follows.  $\rho_1\equiv\rho(770)\mbox{,
}\rho_2\equiv\rho(1450)\mbox{, }\rho_3\equiv\rho(1700);$
$$G=\left(%
\begin{array}{ccc}
  D_{\rho_1} & -\Pi_{\rho_1\rho_2} & -\Pi_{\rho_1\rho_3}  \\
  -\Pi_{\rho_1\rho_2} & D_{\rho_2} & -\Pi_{\rho_2\rho_3}  \\
  -\Pi_{\rho_1\rho_3} & -\Pi_{\rho_2\rho_3} & D_{\rho_3}  \\
  \end{array}%
\right);$$ is the matrix of inverse propagators, $g_{ij}$ are its
matrix elements multiplied by $\Delta={\rm det}G$; $g_{\gamma
V}=m^2_V/g_V$ where $g_V$ enters the leptonic partial widths like
$\Gamma_{V\to e^+e^-}=4\pi\alpha^2m_V/3g^2_V$;
$D_{\rho_i}=m^2_{\rho_i}-s-\Pi_{\rho_i\rho_i}$. The diagonal
polarization operators are $\Pi_{\rho_i\rho_i}=
g^2_{\rho_i\pi\pi}\left[\Pi(s,m^2_{\rho_i},m^2_\pi)+
\frac{1}{2}\Pi(s,m^2_{\rho_i},m^2_K)\right]$, the non-diagonal
ones are expressed through the diagonal one:
\begin{eqnarray}
\Pi_{\rho_1\rho_{2,3}}&=&\frac{g_{\rho_{2,3}\pi\pi}}{g_{\rho_1\pi\pi}}\Pi_{\rho_1\rho_1},\nonumber\\
\Pi_{\rho_2\rho_3}&=&\frac{g_{\rho_2\pi\pi}g_{\rho_3\pi\pi}}{g^2_{\rho_1\pi\pi}}\Pi_{\rho_1\rho_1}
+sa_{23}.\end{eqnarray}The quantity $a_{23}$ is free parameter.
The $\omega(782)$ propagator is
$D_\omega=m^2_\omega-s-i\sqrt{s}\Gamma_\omega(s)$, where
$\Gamma_\omega(s)$ includes the $3\pi$ and radiative decay modes,
$\Pi_{\rho_1\omega}=\frac{s}{m^2_\omega}\Pi_{\rho_1\omega}^\prime+
i\sqrt{s}\left[\frac{1}{3}\Gamma_{\omega\pi\gamma}(s)+
3\Gamma_{\omega\eta\gamma}(s)\right]$ is responsible for the
$\rho(770)-\omega(782)$ mixing.

The expression for the polarization operator of the vector meson
$V$ is obtained from the dispersion representation:
\begin{eqnarray}
\frac{\Pi_{VV}^{(PP)}(s)}{s}&=&\frac{g^2_{VPP}}{6\pi^2}
\int_{4m_P^2}^\infty\frac{q^3_{PP}(s^\prime)}{s^{\prime3/2}(s^\prime-s-i\varepsilon)}ds^\prime
\end{eqnarray}
Remaining logarithmic divergence cancels after subtracting the
real part of the above expression   at $s=m^2_V$. The resulting
expression is represented in the form
$\Pi(s,m^2_V,m^2_P)=\Pi_0+\Pi_1$, where
\begin{eqnarray}
\Pi_0&=&\frac{s}{48\pi^2}\left[8m^2_P\left(\frac{1}{m^2_V}-\frac{1}{s}\right)
+v^3_P(m^2_V)\times\right.\nonumber\\&&\left.
\ln\frac{1+v_P(m^2_V)}{1+v_P(m^2_V)}\theta(m_V-2m_P)
-\right.\nonumber\\&&\left.
2\bar v^3_P(m^2_V)\arctan\frac{1}{\bar v_P}\theta(2m_P-m_V)\right],\nonumber\\
\Pi_1&=&\frac{sv^3_P(s)}{48\pi^2}\left[i\pi-\ln\frac{1+v_P(s)}{1-v_P(s)}\right]
\theta(s-4m^2_P)+\nonumber\\&&2\bar v^3_P(s)\arctan\frac{1}{\bar
v_P(s)}\theta(4m^2_P-s)\theta(s)
-\nonumber\\&&v^3_P(s)\ln\frac{v_P(s)+1}{v_P(s)-1}\theta(-s),
\end{eqnarray}
and $v_P(s)=\sqrt{1-\frac{4m^2_P}{s}}$, $\bar
v_P(s)=\sqrt{\frac{4m^2_P}{s}-1}$, $\theta$ is the step function.
The necessary details of the derivation and parametrization  are
given elsewhere \cite{ach11}.

\begin{figure}
\includegraphics[width=40mm]{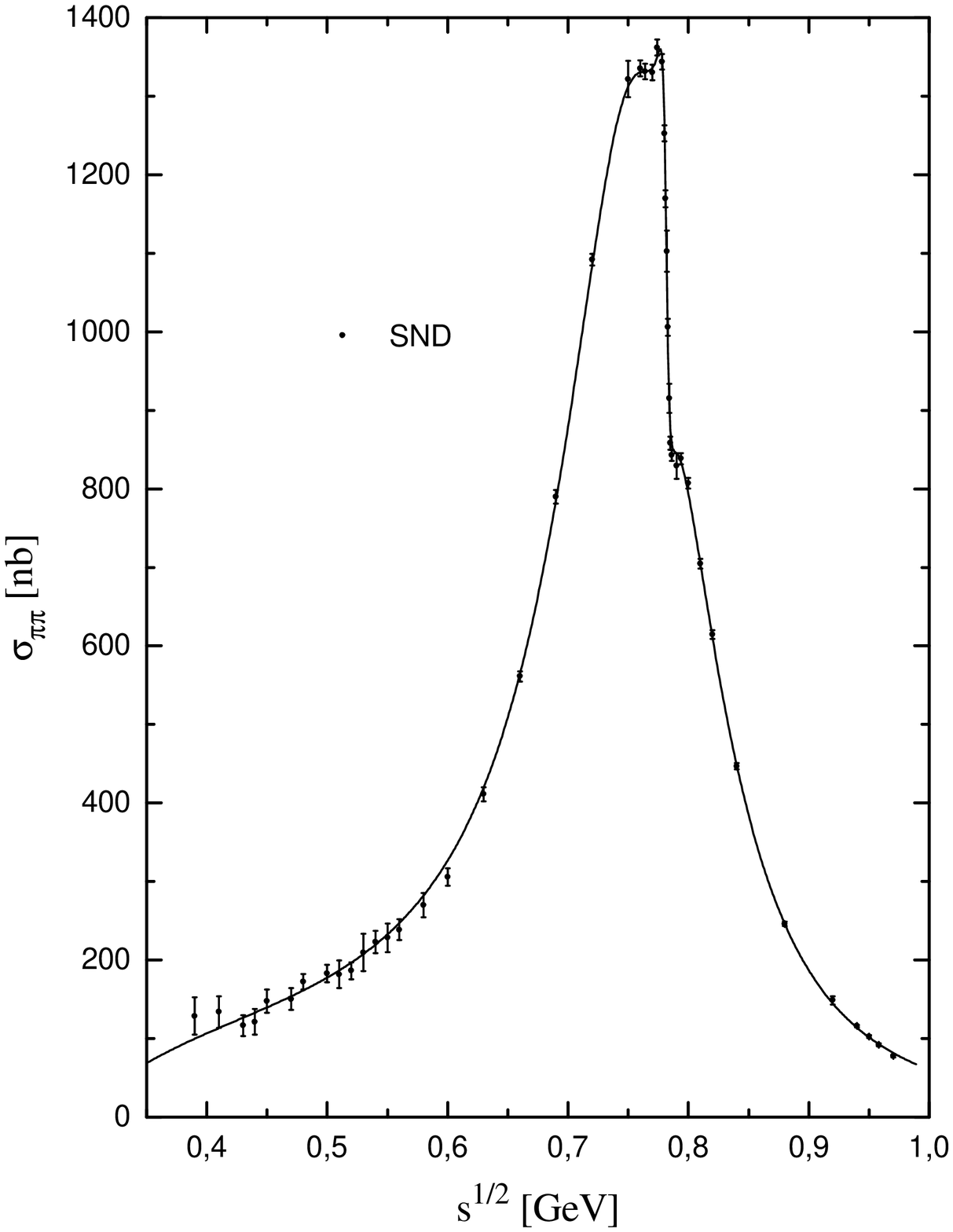}
\caption{\label{spisnd} The bare cross section,
Eq.~(\ref{sigbare}), calculated with the resonance parameters
obtained from fitting the SND data \cite{snd} listed in Table
\ref{table1}. }
\end{figure}

\begin{figure}
\includegraphics[width=40mm]{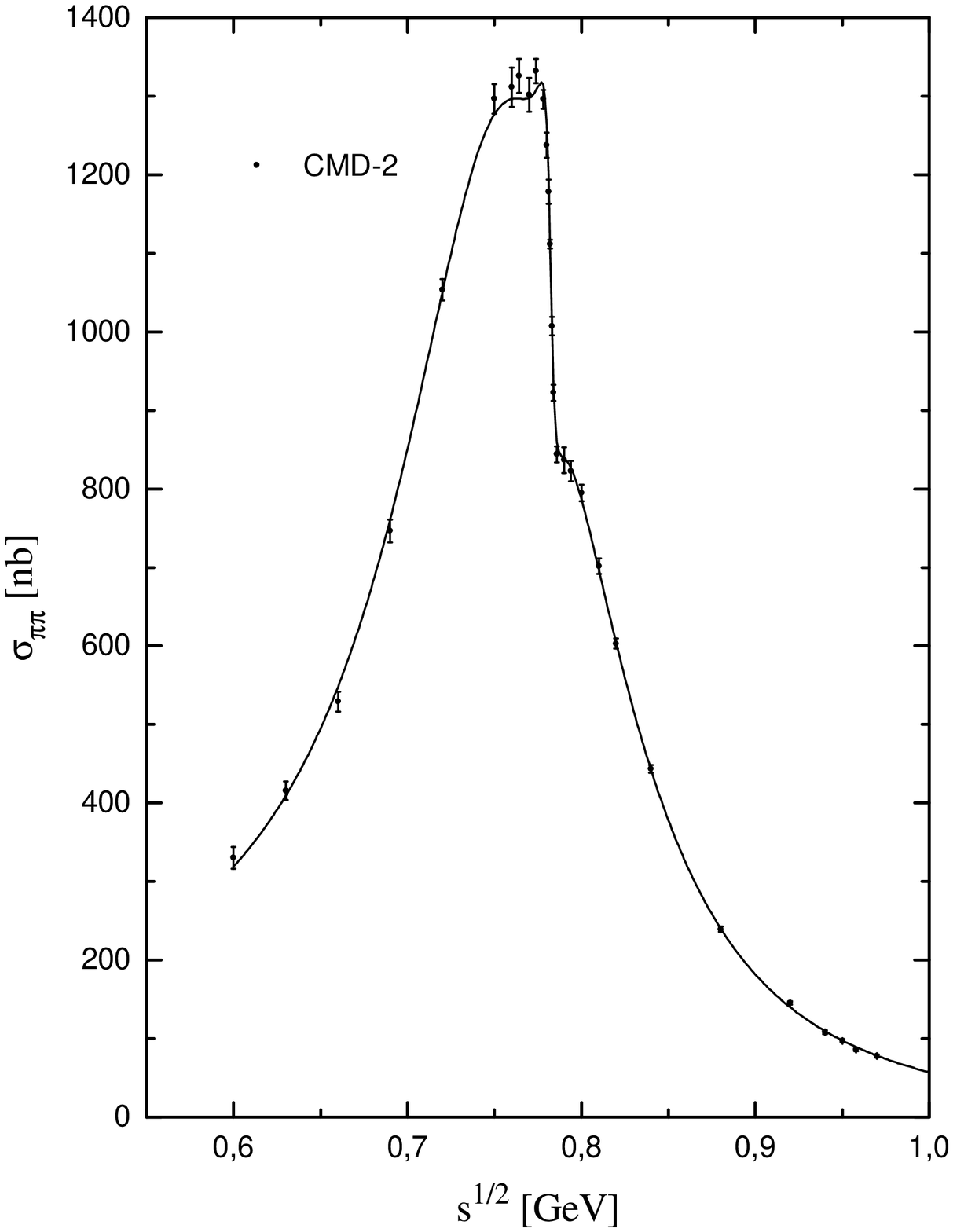}
\caption{\label{spicmd}The same as in Fig.~\ref{spisnd}, but
evaluated with the parameters obtained from fitting the CMD-2 data
\cite{cmd}. }
\end{figure}

\begin{figure}
\includegraphics[width=40mm]{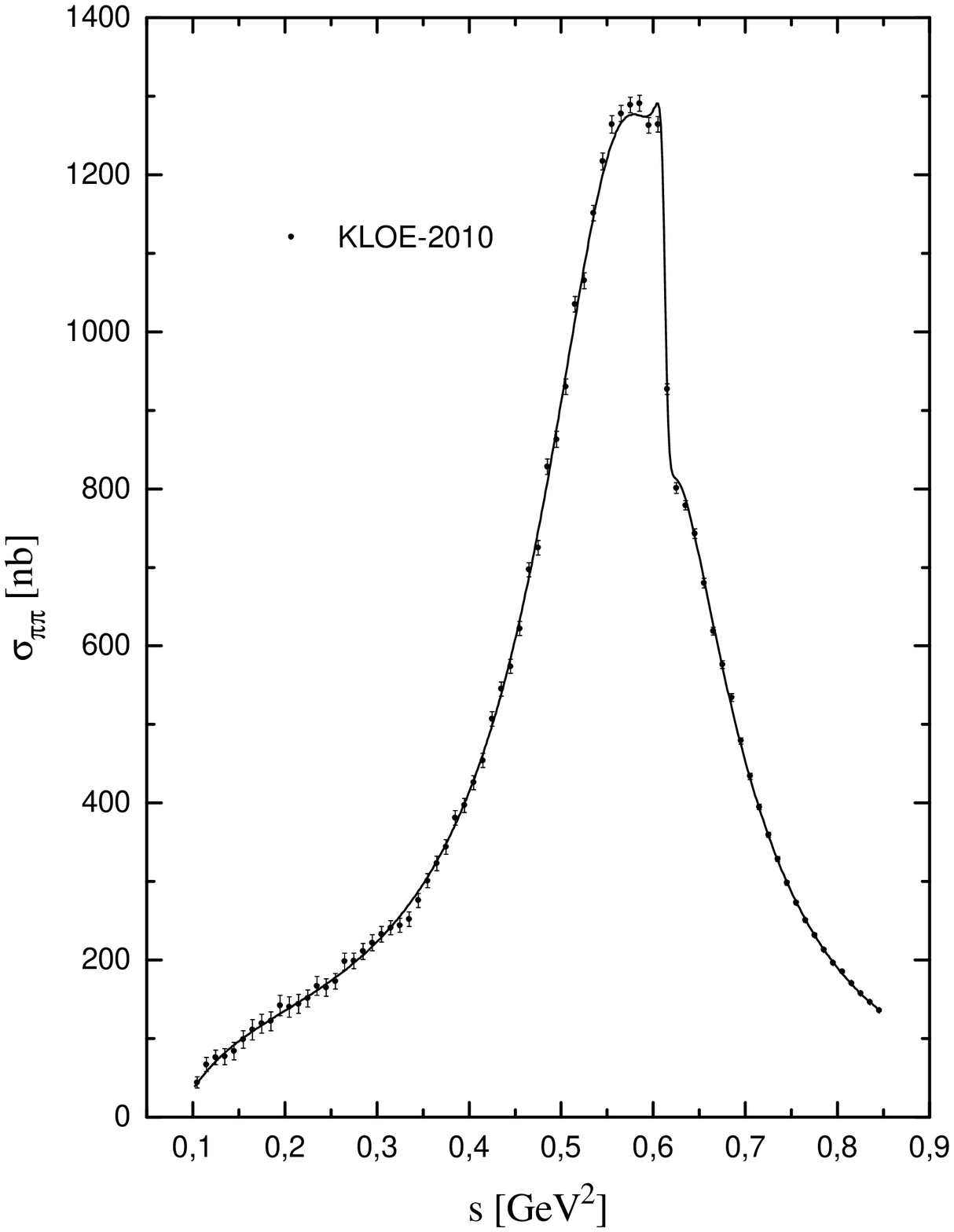}
\caption{\label{spikloe}The same as in Fig.~\ref{spisnd}, but
evaluated with the parameters obtained from fitting the KLOE-2010
data \cite{kloe}. }
\end{figure}

\begin{figure}
\includegraphics[width=40mm]{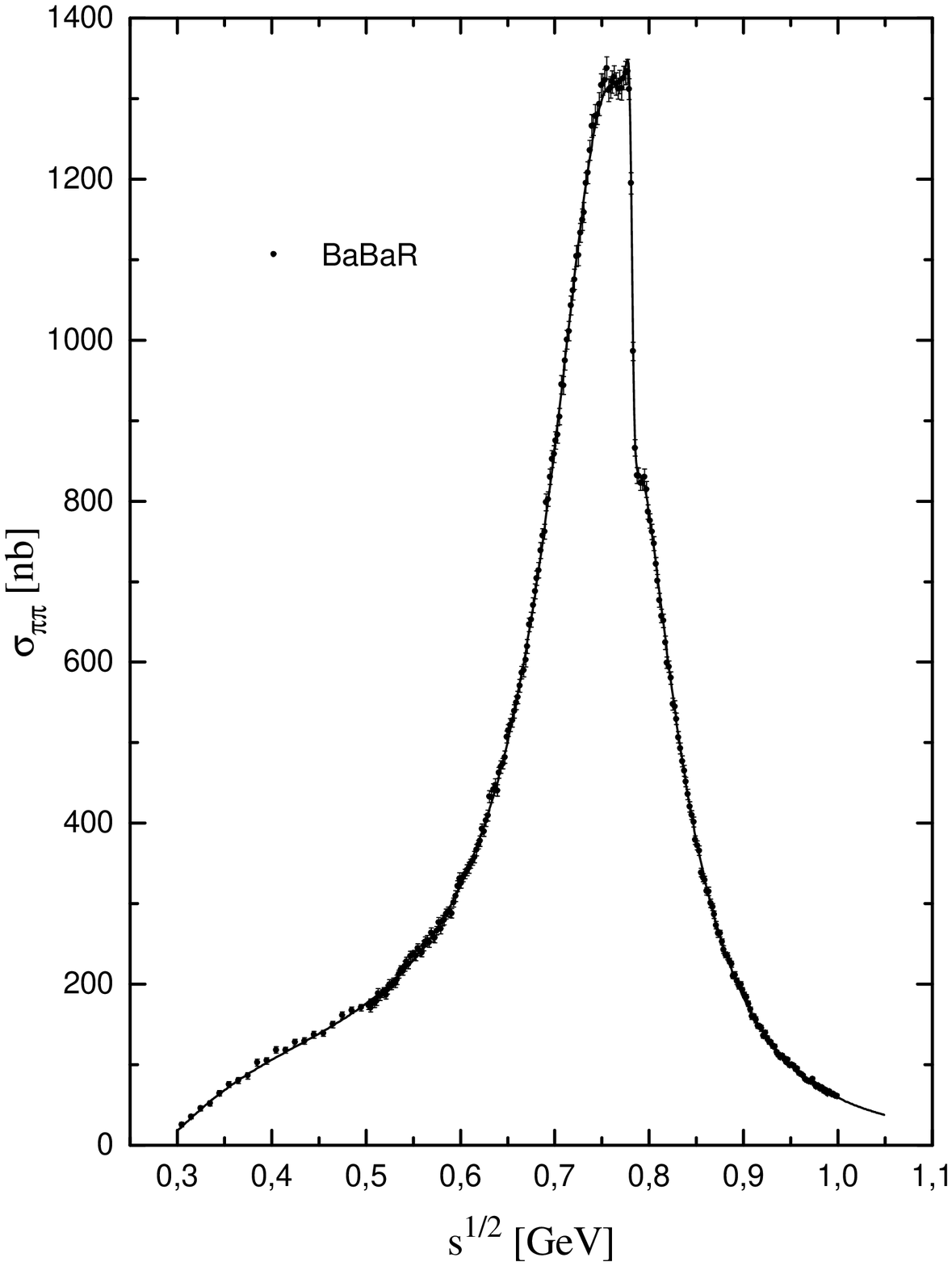}
\caption{\label{spibabar}The same as in Fig.~\ref{spisnd}, but
evaluated with the parameters obtained from fitting the BaBaR data
\cite{babar} restricted to the energies $\sqrt{s}\leq1$ GeV.}
\end{figure}

The quantity to fit is the bare cross section
\begin{equation}
\sigma_{\rm
bare}=\frac{8\pi\alpha^2}{3s^{5/2}}|F_\pi(s)|^2q^3_\pi(s)
\left[1+\frac{\alpha}{\pi}a(s)\right],\label{sigbare}
\end{equation}
where $F_\pi(s)$ is given by Eq.~(\ref{Fpi}),
$q_\pi(s)=\sqrt{s}v_\pi(s)/2$ is the momentum of the final pion,
and the function $a(s)$ allows for the radiation of a photon by
the final pions in the point-like approximation
\cite{schwing,deers,melnik,hoef}. The masses of the heavier vector
mesons a kept fixed: $m_{\rho_2}=1450$ MeV, $m_{\rho_3}=1700$ MeV.
The set of free parameters is $m_{\rho_1}$, $g_{\rho_1\pi\pi}$,
$g_{\rho_1}$, $m_\omega$, $g_\omega$, $\Pi^\prime_{\rho_1\omega}$,
$g_{\rho_2\pi\pi}$, $g_{\rho_2}$, $g_{\rho_3\pi\pi}$, and
$a_{23}$. Their obtained values, found from fitting the bare cross
section Eq.~(\ref{sigbare}), side-by-side with the corresponding
$\chi^2$ per number of degrees of freedom, are listed in  Table
\ref{table1} separately for the four independent measurements of
SND \cite{snd}, CMD-2 \cite{cmd}, KLOE \cite{kloe}, and the BaBaR
data \cite{babar} restricted to the low-energy range
$\sqrt{s}\leq1$ GeV.
\begin{table*}
\caption{\label{table1}The resonance parameters found from fitting
the data SND \cite{snd}, CMD-2 \cite{cmd}, KLOE10 \cite{kloe}, and
the BaBaR data \cite{babar} restricted to the energies
$\sqrt{s}\leq1$ GeV.}
\begin{ruledtabular}
\begin{tabular}{lllll}
 parameter&SND&CMD-2&KLOE10&BaBaR\\ \hline
 $m_{\rho_1}$
 [MeV]&$773.76\pm0.21$&$774.70\pm0.26$&$774.36\pm0.12$&$773.92\pm0.10$\\
 $g_{\rho_1\pi\pi}$&$5.798\pm0.006$&$5.785\pm0.008$&$5.778\pm0.006$&$5.785\pm0.004$\\
 $g_{\rho_1}$&$5.130\pm0.004$&$5.193\pm0.006$&$5.242\pm0.003$&$5.167\pm0.002$\\
 $m_\omega$
 [MeV]&$781.76\pm0.08$&$782.33\pm0.06$&$782.94\pm0.11$&$782.04\pm0.10$\\
 $g_\omega$&$17.13\pm0.30$&$18.43\pm0.47$&$18.27\pm0.45$&$17.05\pm0.29$\\
 $10^3\Pi^\prime_{\rho_1\omega}$
 [GeV$^2$]&$4.00\pm0.07$&$3.97\pm0.10$&$3.98\pm0.09$&$4.00\pm0.06$\\
 $g_{\rho_2\pi\pi}$&$0.71\pm0.35$&$0.79\pm0.26$&$0.019\pm0.004$&$0.21\pm0.04$\\
 $g_{\rho_2}$&$8.0\pm4.4$&$7.6\pm3.4$&$0.22\pm0.07$&$4.0\pm1.0$\\
 $g_{\rho_3\pi\pi}$&$0.20^{+1.20}_{-0.17}$&$0.76\pm0.75$&$0.055^{+0.088}_{-0.043}$&$0.011^{+0.479}_{-0.007}$\\
 $a_{23}$&$0.002\pm0.011$&$-0.016\pm0.057$&$-0.014\pm0.040$&$-0.0005\pm0.0009$\\
 $\chi^2/N_{\rm d.o.f.}$&54/35&34/19&87/65&216/260\\
 $r_\pi$[fm]&$0.635\pm0.054$&$0.646\pm0.059$&$0.668\pm0.039$&$0.668\pm0.053$\\
\end{tabular}
\end{ruledtabular}
\end{table*}
The bare cross section evaluated with the parameters of Table
\ref{table1} is compared with the SND \cite{snd}, CMD-2
\cite{cmd}, KLOE \cite{kloe}, and BaBaR \cite{babar} data  shown
in Figs.~\ref{spisnd}, \ref{spicmd}, \ref{spikloe}, and
\ref{spibabar}, respectively. The bottom line of this Table shows
the values of the pion charge radius, $
r_\pi=\sqrt{6\frac{dF_\pi(s)}{ds}}|_{s\to0}$, calculated with the
resonance parameters listed in the Table. For comparison, the
averaged value of the pion charge radius cited by the PDG
\cite{pdg} is $r_\pi=0.672\pm0.008$ fm.

\begin{figure}
\includegraphics[width=40mm]{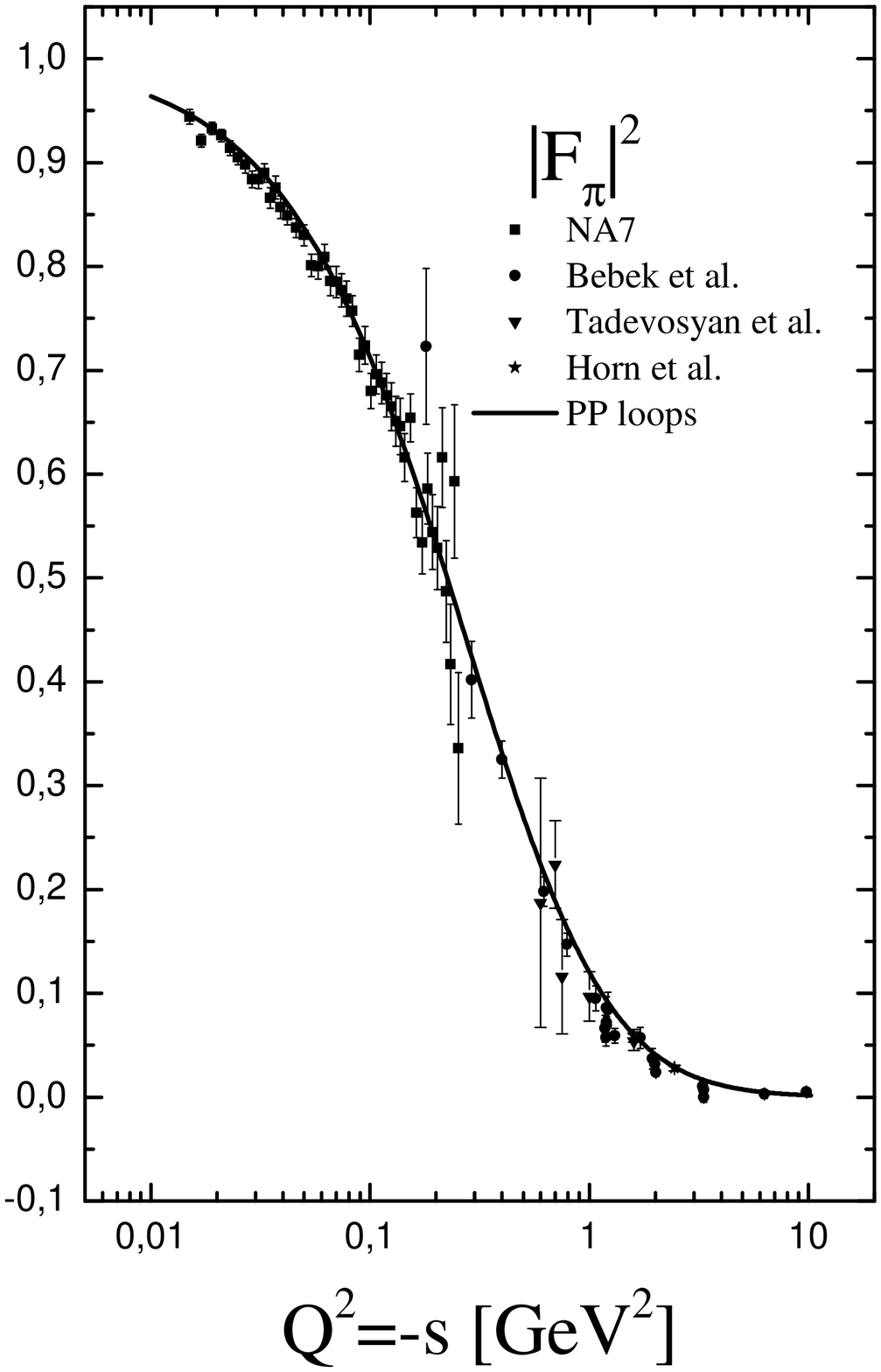}
\caption{\label{euclid}The pion form factor squared in the
space-like region $s<0$ evaluated using the resonance parameters
of the Table \ref{table1}, the BaBaR column. The experimental data
are: NA7 \cite{amendolia}, Bebek {\it et al.} \cite{bebek}, Horn
{\it et al.} \cite{horn}, Tadevosyan {\it et al.}\cite{tadev}.}
\end{figure}

An important check of the expression for the pion form factor
Eq.~(\ref{Fpi}) and the consistency of the fits is the
continuation to the space-like region $s<0$ accessible in the
scattering processes. To this end, one should   take the
expression for $F_\pi(s)$ at  $s<0$. Having in mind that the
$\rho(770)-\omega(782)$ mixing in the region $s<0$ is negligibly
small one can calculate $F_\pi(s)$ in this region. The results are
shown in Fig.~\ref{euclid}, where the comparison with the data
\cite{amendolia,bebek,horn,tadev}  is presented in the case of the
resonance parameters found from fitting the BaBaR data
\cite{babar}. As for the curves corresponding to the parameters
found from fitting SND, CMD-2, and KLOE data, they coincide,
within the errors, with the curve shown and hence are not drawn in
Fig.~\ref{euclid}. We emphasize that  the data
\cite{amendolia,bebek,horn,tadev} are not included to the fits.
Hence, a good agreement, demonstrated in Figs.~\ref{euclid} makes
the evidence in favor of the validity of Eq.~(\ref{Fpi}) for the
pion form factor.

\begin{figure}
\includegraphics[width=40mm]{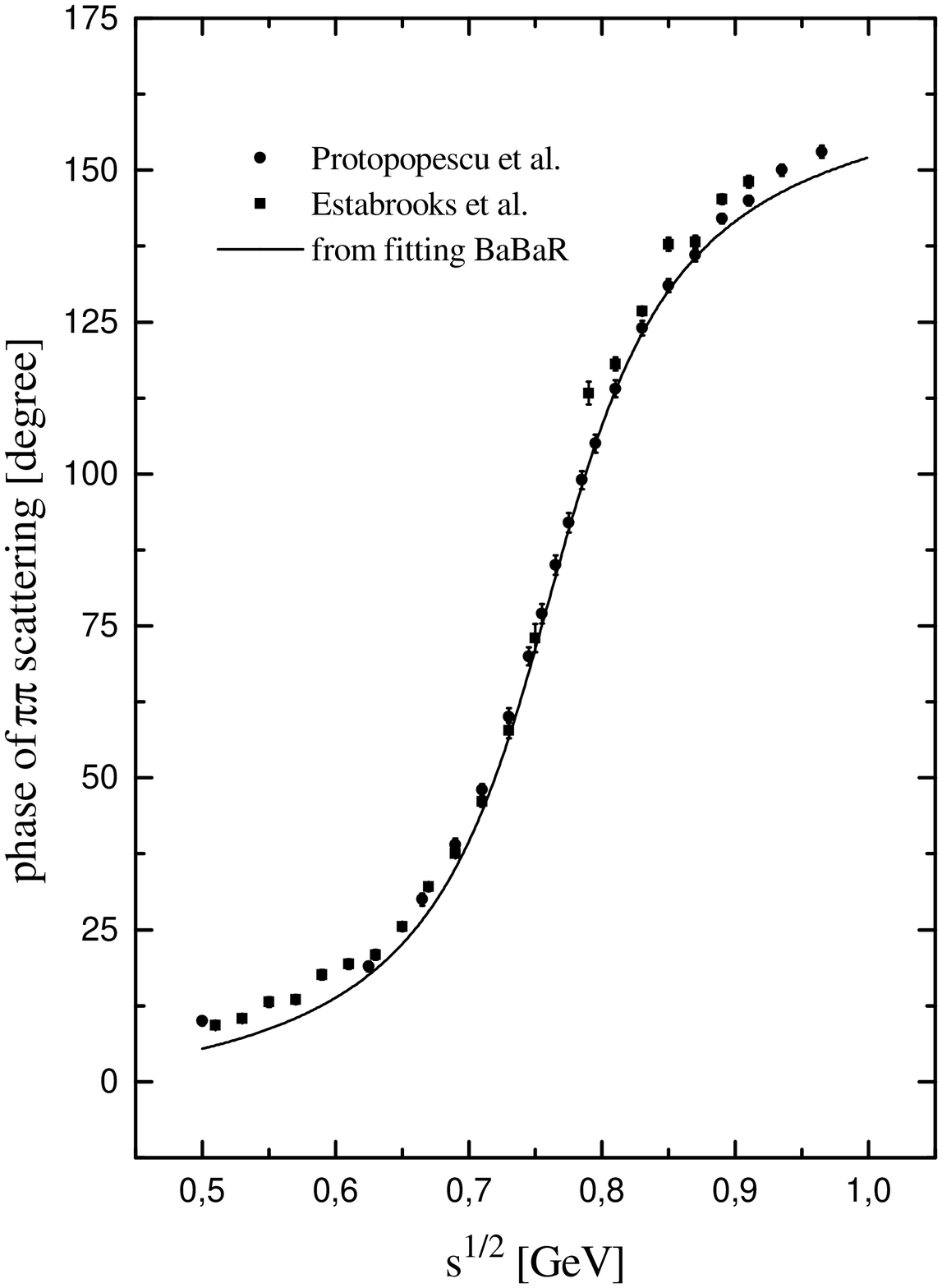}
\caption{\label{phase}The phase shift $\delta^1_1$ of $\pi\pi$
scattering. The data are, respectively, Protopopescu et al.
\cite{proto} and Estabrooks et al. \cite{estab}. The curves
corresponding to the parameters obtained from fitting the SND,
CMD-2, and KLOE data are not shown because they coincide with the
curve evaluated using the parameters from the fit of the BaBaR
data, shown here.}
\end{figure}

Note that the above treatment does not require the commonly
accepted Blatt -- Weisskopf centrifugal factor $(1+R^2_\pi
k^2_R)/(1+R^2_\pi k^2)$, where $k$ is the pion momentum, in the
expression for $\Gamma_{\rho\pi\pi}(s)$ \cite{pdg}. The fact is
that the usage of $R_\pi$ dependent centrifugal barrier
penetration factor in particle physics (for example, in the case
of the $\rho(770)$ meson \cite{pdg}), results in the problem which
is overlooked. Indeed, the meaning of $R_\pi$ is that this
quantity is the characteristic of the potential (or the
$t$-channel exchange in field theory) resulting in the phase
$\delta_{\rm bg}$ of the potential  scattering in addition to the
resonance phase \cite{blatt}. For example, in case of the $P$-wave
scattering in the potential $U(r)=G\delta(r-R_\pi)$ where the
resonance scattering is possible, the background phase is
$\delta_{\rm bg}=-R_\pi k+\arctan(R_\pi k)$.  At the usual value
of $R_\pi\sim1$ fm, $\delta_{\rm bg}$ is not small. However, in
the $\rho$ meson region, the background phase shift $\delta_{\rm
bg}$ is negligible and the phase shift $\delta^1_1$ is completely
determined by the resonance. See Fig.~\ref{phase}, where shown are
the phase $\delta^1_1$ calculated with the parameters from the
BaBaR column of the table \ref{table1} and the data points from
Refs.~\cite{proto,estab} is presented. Therefore, the descriptions
of the hadronic resonance distributions taking into account the
parameter $R_\pi$ have a dubious character.

To conclude, the new expression, Eq.~(\ref{Fpi}), for the pion
form factor $F_\pi(s)$ is obtained which  gives a good description
of the data of SND, CMD-2, KLOE, BaBaR on $\pi^+\pi^-$ production
in $e^+e^-$ annihilation at $\sqrt{s}<1$ GeV,  describes the
scattering kinematical domain, and   does not contradict the data
on $\pi\pi$ scattering phase $\delta^1_1$. Going to higher
energies demands the inclusion of the vector -- pseudoscalar and
axial-vector -- pseudoscalar loops.  This work is now at progress.

We are grateful to M.~N.~Achasov for numerous discussions which
stimulated the present work. The  work is supported in part by
Russian Foundation for Basic Research Grant no. RFFI-10-02-00016
and the Interdisciplinary project No 102 of Siberian Division of
Russian Academy of Sciences.

\end{document}